\documentclass[useAMS,usenatbib]{mnras}
\usepackage{amssymb}
\usepackage{deluxetable}

\title[VLBI images at 327 MHz of CSS/GPS sources]{VLBI images at 327
  MHz of compact steep spectrum and GHz-peaked spectrum sources from the 3C and PW samples}
\author[D. Dallacasa et al. ]
  {D. Dallacasa$^{1,2}$\thanks{E-mail: ddallaca@ira.inaf.it},
M. Orienti$^{2}$, C. Fanti$^{2}$, R. Fanti$^{2}$\\
$^1$Dipartimento di Astronomia, Universit\`a di Bologna, via Ranzani 1,
I-40127, Bologna, Italy \\
$^2$INAF -- Istituto di Radioastronomia, via Gobetti 101, I-40129, Bologna,
Italy  }
\date{Received \today; accepted ?}

\pagerange{\pageref{firstpage}--\pageref{lastpage}} \pubyear{2002}

\def\LaTeX{L\kern-.36em\raise.3ex\hbox{a}\kern-.15em
    T\kern-.1667em\lower.7ex\hbox{E}\kern-.125emX}

\begin{document}

\label{firstpage}

\maketitle

\begin{abstract}
We present results on global very long baseline interferometry (VLBI) 
observations at 327 MHz of eighteen compact
steep-spectrum (CSS) and GHz-peaked spectrum (GPS) radio sources from the 3C and
the Peacock \& Wall  
catalogues. About 80 per cent of the sources have a 'double/triple'
structure. The radio emission at 327 MHz is dominated by
  steep-spectrum extended
  structures, while compact regions become predominant at higher
  frequencies. As a consequence, we could unambiguously detect the core region only
  in three sources, likely due to self-absorption affecting its
  emission at this low frequency. Despite their low surface brightness, lobes store
  the majority of the source energy budget, whose correct estimate
  is a key ingredient in tackling the radio source
  evolution. Low-frequency VLBI observations able to disentangle the
  lobe emission from that of other regions are therefore the best way
  to infer the energetics of these objects.
Dynamical ages estimated from energy budget arguments provide values
between 2$\times$10$^{3}$ and 5$\times$10$^{4}$ yr, in 
agreement with the radiative ages estimated from the fit of the
integrated synchrotron spectrum, further supporting the youth of these
objects. A discrepancy between radiative and dynamical ages
  is observed in a few sources where the integrated spectrum is
  dominated by hotspots. In this case the radiative age likely
  represents the time spent by the particles in these regions,
  rather than the source age.

\end{abstract}


\section{Introduction}

Powerful radio sources hosted in elliptical galaxies and
quasars represent about 10 per cent of the population of active
galactic nuclei. Although
their linear sizes span several orders of
magnitude, from a few parsecs up to several Mpc, the radio morphology is
described by the same ingredients: a central core, where the
relativistic plasma is produced and accelerated; 
two bipolar jets that channel the plasma
towards the outermost regions; the hotspots, that mark the place where
the jet impacts with the external medium;
and the lobes where particles are deposited after being
further accelerated in the hotspot, and where eventually they age.
As a consequence
of the dominant mechanism (i.e. acceleration/cooling) the spectral
shape of the various components is different. For example the core is
characterized by a flat or inverted spectrum, indicating the presence
of synchrotron self-absorption, and more properly should be defined as
the region where the jet becomes optically thick at a given frequency. 
The spectrum of the hotspots is
usually a power-law with spectral index $\alpha \sim$ 0.5--0.7
($S_{\nu} \propto \nu^{- \alpha}$) marking the presence of high Mach number
shock-induced particle acceleration, while the lobes have a steep
spectrum with $\alpha > 0.7$, indicating significant energy losses.\\
\indent Flux-limited catalogues selected at low frequencies, like the 3CRR \citep{laing83}, 
are dominated by steep-spectrum
radio sources, usually hosted in galaxies, 
whose radio emission mainly arises from the lobes. On
the other hand, in catalogues selected at high frequencies, like the
Peacock \& Wall \citep{pw81} and the Australia Telescope 20-GHz Survey
\citep[AT20G][]{murphy10}, we 
expect a larger fraction of sources with bright and compact cores
typically with a flat spectrum,
and usually associated with quasars. Interestingly, in both low- and
high-frequency selected catalogues about 10-15 per cent of the sources
are unresolved on arcsecond scales. Statistical studies showed that
the majority of these objects are intrinsically 
compact with linear sizes $< 20$ kpc, 
and not foreshortened by projection effects \citep{fanti90}. When
observed with adequate spatial resolution they usually show a
two-sided radio structure resembling a scaled-down version of
Fanaroff-Riley radio sources. Depending on their linear size (LS) they are
termed compact symmetric objects (CSO, LS$<$ 1 kpc) or medium-sized
symmetric objects (MSO, 1 kpc $<$LS$<$ 20 kpc).
The main characteristic of CSO/MSO sources is the steep synchrotron radio spectrum
that turns over at frequencies between $\sim$50 MHz and a few
GHz. Depending on the peak frequency $\nu_{\rm p}$, these compact objects are termed
compact steep-spectrum (CSS, $\nu_{\rm p} <$500 MHz) and
gigahertz-peaked spectrum (GPS, $\nu_{\rm p} >$ 500 MHz) radio
sources \citep{odea98}. CSS/GPS sources are optically
identified with both galaxies and quasars, the latter being more
commonly found among high-frequency peaking GPS sources \citep[e.g.][]{torniainen07}. \\
\indent Many studies were carried out to understand the nature of CSS/GPS
sources, i.e. whether they are small because in an early stage of
their evolution \citep[e.g.][]{fanti95}, or frustrated by an
extraordinary dense environment \citep[e.g.][]{vanbreugel84}. Estimates of
kinematic and radiative ages of a few thousand years 
\citep{polatidis03,murgia03,gugliucci05,giroletti09,an12a}, together with the lack
of any evidence of uncommonly rich ambient medium
\citep[e.g.][]{fanti00,siemiginowska05}, strongly support the genuine youth of
these objects. However, in some sources there is evidence of
jet-medium interaction which may temporarily frustrate the source
expansion
\citep[e.g.][]{labiano06,dd13,morganti13,sobolewska19,zovaro19}. 
Detailed information on GPS/CSS sources can be found in the 
  review by \citet{odea20}.\\
\indent Evolutionary models have been proposed to trace the
various stages of the source growth
\citep[e.g][]{fanti95,readhead96,snellen00,alexander00,an12} and to determine how
the physical parameters evolve. The main ingredient at the basis of the
source evolution is the balance between the jet power and the ram
pressure on the external medium. Therefore, a correct estimate of the
jet thrust and the energetics of the radio source is fundamental for
drawing a clear picture of the source evolution. \\
\indent In this paper we present results of global
very long baseline interferometry (VLBI) observations at 327 MHz of 18
CSS/GPS objects from the 3C and Peacock \& Wall catalogues. 
Our aim is the characterization of the radio morphology
at low frequency and the estimate of the energetics and
dynamical age of the radio sources. 
To achieve this goal we need observations at low frequency 
to detect the steep-spectrum emission from the lobes, where
the majority of energy is stored, and with high spatial resolution to
disentangle the lobe emission from that of other regions. 
For CSS/GPS have compact linear size, 
only low-frequency global VLBI observations
can fulfill the requirements. So far only a handful of works based on VLBI images at 327
MHz are available. They mainly focus on variable extragalactic
sources \citep{altschuler95,chuprikov99,rampadarath09} or on deep wide
field VLBI survey \citep{lenc08}. 
The data
presented here are the deepest and with the highest angular resolution
at 327 MHz for a sample of CSS/GPS published so far. \\
 \indent This paper is organized as follows: in Section 2
we describe the radio observations and data reduction; results are
reported in Section 3 and
discussed in Section 4. A brief summary is presented in Section 5.\\
Throughout this paper, we assume 
$H_{0} = 71$ km s$^{-1}$ Mpc$^{-1}$, $\Omega_{\rm M} = 0.27$,
$\Omega_{\Lambda} = 0.73$, in a flat Universe. The spectral index is
defined as $S$($\nu$) $\propto \nu^{- \alpha}$. \\

\begin{table*}
\caption{Antennas participating the observations. Column 1: observing
  date; Column 2: observing mode; Column 3: antennas. Wb=Westerbork,
  Jb=Jodrell Bank, Nt=Noto, Tr=Torun, Sm=Simeiz (Crimea), Gb=43-meter
  Green Bank telescope, Pt=Pie Town,
  La=Los Alamos, Nl=North Liberty Br=Brewster, Ov=Owens Valley,
  Hn=Hancock, VLA=Very Large Array - single antenna, Kp=Kitt Peak, Mk=Mauna Kea, Fd=Fort Davis, Sc=St. Croix.}
\begin{center}
\begin{tabular}{ccc}
\hline
Date & Obs. mode & Antennas \\
\hline
04-11-1992 & MkII & Wb, Jb, Nt, Tr, Sm, Gb, Pt, La, Nl, Br, Ov,
Hn, VLA\\
20-02-1995 &MkIII & Wb, Jb, Nt, Gb, Pt, La, Nl, Br, Hn, Kp, Mk, Fd, Sc\\
\hline
\end{tabular}
\end{center} 
\label{antenna}
\end{table*}

\section{Observations and Data Reduction}

\begin{figure}
\begin{center}
\includegraphics{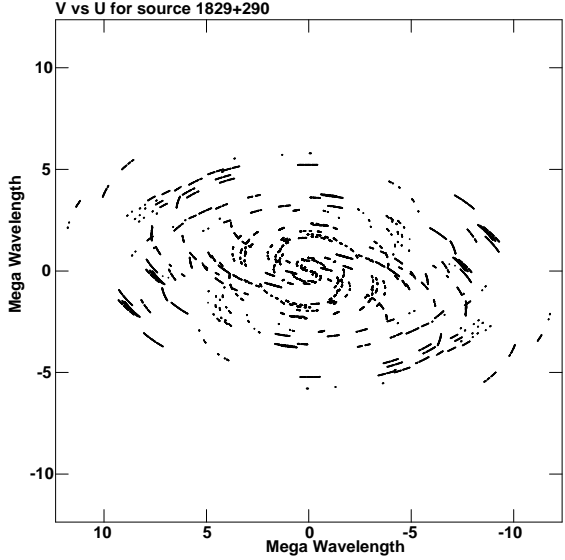}
\vspace{8.0cm}
\caption{({\it u,v}) coverage of the VLBI network at 327 MHz in units of wavelengths
  for the source 1829$+$290.}
\label{uv}
\end{center}
\end{figure}

The observations presented in this paper were obtained with VLBI
networks at 327 MHz in different observing modes (MkII and MkIII) on 1992
November 4 and 1995 February 20. Antennas involved
in the observing runs are listed in Table \ref{antenna}. Observations
were carried out in snapshot mode in single polarization with a total
bandwidth of 2 MHz and 14 MHz for MkII and MkIII, respectively. Correlation
was performed at the Caltech-JPL VLBI Correlator. A-priori amplitude
calibration was made using the system temperature measurements
and the antenna gains for each telescope. The uncertainty on the
flux density scale is approximately 10 per cent.
The target sources
and the calibrators were fringe-fitted with a solution interval of 60
seconds. The coherent time at this frequency is rather short ($\sim$10
seconds). Therefore, the 2-sec integration data underwent a number of
self-calibration iteration before time averaging.
Editing, fringe-fitting, and all the usual operations on VLBI data sets
were performed using the Astronomical Image Processing System (\texttt{AIPS})
software. Final images were produced after a few imaging and
phase-only self-calibration iterations. 
The final noise on the image
plane is typically between 1.5 and 5 mJy beam$^{-1}$. 
In Fig. \ref{uv} we show the ({\it u,v}) coverage for the source
1829$+$290 as an example. We note that the lack of baselines shorter
than 0.1 M$\lambda$ prevents the detection of regions larger than
$\sim$2 arcseconds, i.e. comparable to the lobe emission detected by
MERLIN observations in a few sources (see Section 3).\\

\begin{table*}
\caption{CSS/GPS sources from the 3C and PW samples. Column 1: source
  name; Column 2: optical identification: G=galaxy, Q=quasar; Column 3: 
  redshift; Column 4: peak frequency from \citet{fanti90}; 
  Column 5: total flux density at 327 MHz from the VLBI data
  presented in this paper (fraction of flux density that is
    recovered in VLBI images);
  Column 6: total flux density at 327 MHz from
  observations with low spatial resolution; Columns 7 and 8: largest angular
  size and largest linear size derived from VLBI images presented 
in this paper; Column 9: reference for the low-resolution flux
density. 1: \citet{jeyakumar00}, 2: WENSS \citep{rengelink97}, 
3: \citet{kuhr81}, 4:
\citet{douglas96}. Column 10: source morphology as derived from our
VLBI images at 327 MHz. C=complex morphology, D=double structure,
T=triple structure.}
\begin{center}
\begin{tabular}{cccccccccc}
\hline
Source   & Id& z    &$\nu_{\rm p}$&$S_{\rm VLBI}^{\rm 92\;cm}$&$S_{\rm tot}^{\rm 92\;cm}$&  LAS &  LLS&Ref&Morph.\\
        &   & & MHz     &   Jy   (per cent)     &   Jy     &arcsec&  kpc&    & \\
(1)&(2)&(3)&(4)&(5)&(6)&(7)&(8)&(9)&(10)\\
\hline
&&&&&&&&&\\
3C\,43   & Q & 1.459& $<20$ &      5.1 (60)  &  8.4     & 1.24 & 10.57& 1 &C\\
3C\,49   & G & 0.621& 120   &      7.4 (97)  &  7.6     & 1.00 &  6.78& 1 &D\\
3C\,93.1 & G & 0.243& 60    &      2.2 (30)  &  7.5     & 0.17 &  0.64& 2 &C\\
3C\,119  & G & 1.023& 150   &     17.9 (100) & 17.7     & 0.42 &  3.39& 2 &C\\
3C\,138  & Q & 0.759& 100   &     19.3 (97)  & 19.8     & 0.62 &  4.57& 1 &T\\
3C\,237  & G & 0.877& 50    &     17.1 (98)  & 17.4\tablenotemark{a}&1.21
&  9.4&  3 &D\\
3C\,241  & G & 1.617& 40    &      5.9 (78)  &  7.6     & 0.83 &  7.10& 1 &D\\
3C\,298  & Q & 1.437& 80    &     20.1 (61)  & 32.9     & 1.49 & 12.69& 1 &T\\
3C\,318  & Q & 1.574& $<40$ &      5.5 (60)  &  9.2\tablenotemark{a}& 1.11
&  9.50& 4 &T\\
3C\,343  & Q & 0.988& 250   &     12.8 (95)  & 13.6     & 0.40 &  0.32& 2 &C\\
3C\,343.1& G & 0.75 & 250   &     11.7 (87) & 13.5     & 0.35 &  2.57&2 &D\\
\hline
&&&&&&&&&\\
0223+341& Q & 2.91  & 250   &     2.2  (60)  &  3.7     & 0.82 &  6.47& 2 &T\\
0316+161& G & 0.907 & 900   &     5.9  (79)  &  7.5\tablenotemark{a}& 0.40
&  3.13& 3 &D\\
0404+768& G & 0.5985& 600   &     8.9  (96)  &  9.3     & 0.130 & 0.87&2 &D\\
1153+317& Q & 0.417 & 100   &     5.4  (65)  &  8.2     & 0.90 &  4.93& 2 &T\\
1819+396& G & 0.798 & 100   &     5.8  (82)  &  7.0     & 0.80 &  6.01& 2 &D\\
1829+290& G & 0.842 & 100   &     3.8  (69)  &  5.5     & 0.38 &  2.93& 2 &T\\
2342+821& Q & 0.735 & 400   &     4.4  (79)  &  5.6     & 0.17 &  1.24& 2 &T\\
\hline
\tablenotetext{a}{The low resolution flux density refers to
  observations at 365 MHz. }
\end{tabular}
\end{center}
\label{source}
\end{table*}

\section{Results}

All the sources are clearly detected in our 327-MHz VLBI observations.
The flux density and the angular size of each
component are measured using the task \texttt{JMFIT} in \texttt{AIPS} 
which performs
a Gaussian fit to the source components on the image plane. In a few
sources multiple components were fitted simultaneously and the
residuals were inspected in order to be sure that no additional
components were necessary.
In the case of diffuse emission
that cannot be fitted by a Gaussian profile, the flux density was
derived using
\texttt{TVSTAT} in \texttt{AIPS} 
which allows the selection of a polygonal region on the
image plane. In this case the angular size was measured on the contour
image and corresponds to roughly twice
the size of the full width half maximum (FWHM)  of a conventional
Gaussian covering a similar area \citep{readhead94}. \\
Often the flux density that is recovered by our VLBI images is only a
fraction of the emission obtained from observations with lower
resolution at the same frequency. Such missing flux density may be due
to additional emission from lobes on large scales that could not be
fully sampled by our VLBI observations (Table \ref{source}).\\
Observational parameters of each source component are reported in
Table \ref{parameter}, whereas VLBI images at 327 MHz are presented in
Figs \ref{3c43}-\ref{2342}. On each image we provide the source name,
the peak flux density (mJy beam$^{-1}$) and the first contour (f.c.)
intensity (mJy beam$^{-1}$), which is three times the off-source noise
level. Contour levels increase by a factor of 2. The restoring beam is
plotted on the bottom left-hand corner.\\

\subsection{Low-frequency radio structure}

\begin{figure}
\begin{center}
\includegraphics{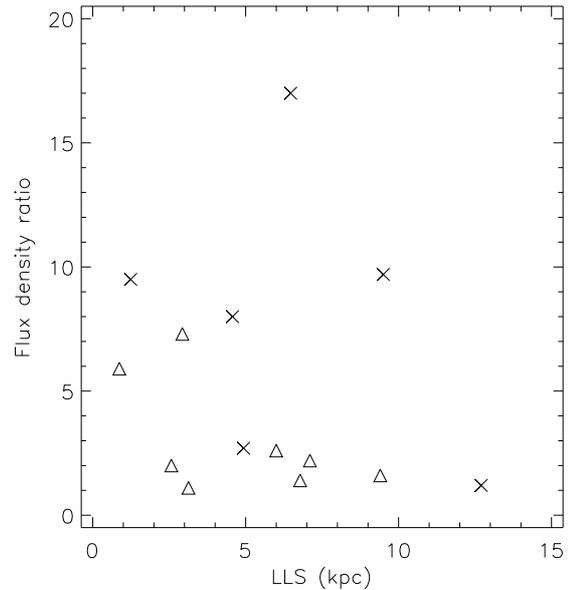}
\vspace{8.5cm}
\caption{Flux density ratio vs largest linear size (LLS) for the 14
  sources with a double/triple morphology. Crosses are quasars, while
  triangles are galaxies.}
\label{lls-flux}
\end{center}
\end{figure}

The structure of all the 18 sources is resolved. 
7 sources have a double morphology, 7 have a triple
structure (although the central component is not always associated with
the core, on the basis of the steepness of the spectrum), 
while the remaining 4 have a complex morphology. 
In 6 sources we could disentangle the emission of at
least one of the lobes from the hotspot contribution, crucial
information for estimate the dynamical age of the radio source (see
Section \ref{discussion}). In 11 sources the flux density recovered
from our observations is $\leq$80 per cent of that measured in
low-resolution observations at the same frequency. This missing flux
is likely caused by the lack of short baselines and poor ({\it u,v})
coverage that make our observations unsuitable for detecting
arcsecod-scale diffuse emission. \\
The core component is detected in 3 sources (3C\,138,
3C\,298, and 0223$+$341), while in 3C\,318 it is tentatively
associated with the central region. The small percentage of core
detection may be explained by either synchrotron self-absorption of
the component at such a low frequency, or a blending of the source
core and emission from other steep-spectrum components
that cannot be resolved by the angular resolution of our
observations.\\
Significant flux density ratio ($R>2$) between the two sides of the radio
sources is found for 10 out of the 14 sources with double/triple
morphology (Fig. \ref{lls-flux}). The largest values (R$>$8) are found in
quasars, and may be ascribed to beaming effects. In general galaxies
have low flux density ratios, with the exception of 0404$+$768, which
is among the smallest of our sample. This is in agreement with what
was found by
\citet{mo16} where the smaller sources are also those with the higher
flux density ratios. The other galaxy with high flux density ratio is
1829$+$290. However, in this source we are considering two internal
components, rather than the lobes which extend more than one arcsec
away from the central region and are not imaged by our
observations. If we consider their flux density from MERLIN
observations we end up with $R \sim 1$ and an LLS of about 20 kpc
\citep{spencer89}, consistent with the fact that asymmetries get less
important as the source grows. \\

\subsection{Spectral indices}

\indent We complemented our 327 MHz VLBI observations with information from the
literature in order to have a more complete description 
of the source structure and spectral information. 
In many cases high resolution data exist in
the literature to allow the analysis of the component spectra from 327
MHz up to 22 GHz. \\
In Table \ref{parameter} we give a low frequency spectral index
($\alpha_{l}$), usually
computed between 0.327 and 0.610 (or 1.7) GHz, and a high frequency
spectral index ($\alpha_{h}$), computed in a range which depends on
the
available data and on the spectral shape (see notes to Table \ref{parameter}).
Unless mentioned otherwise, 
the 610 MHz data used for computing the
spectral index are from 
\citet{nan91}, those at 1.7 and 5 GHz are from
\citet{dd95}, \citet{ludke98}, and \citet{dd13}, while those at 8.4,
15, and 22 GHz are from \citet{mantovani13}. \\
\indent Uncertainty in the spectral index may be large
depending on the component structure: compact regions have small
spectral index errors ($\pm$0.1, mainly from the amplitude
calibration uncertainty), while extended low-surface brightness
components may result artificially steeper/flatter due to missing flux
density caused by the lack of short spacing at the various
frequencies (see Section \ref{section:notes}). When there is no
secure identification of the same component at different frequencies,
we do not provide any value for the spectral index in Table \ref{parameter}.\\

We computed the average high frequency spectral index $\langle \alpha_{h} \rangle$,
the average low frequency spectral index $\langle \alpha_{l} \rangle$, and their
standard deviation for the central components and for the lobes
(inclusive of hotspot emission which cannot be disentangled). For the former we find:

\begin{itemize}

\item $\langle \alpha_{h} \rangle = 0.7 \pm 0.1$, $\sigma=0.2$,
\item $\langle \alpha_{l} \rangle = 0.1 \pm 0.1$, $\sigma=0.4$,

\end{itemize}

\noindent while for the latter we have:\\

\begin{itemize}

\item $\langle \alpha_{h} \rangle = 1.2 \pm 0.1$, $\sigma=0.4$,
\item $\langle \alpha_{l} \rangle = 0.6 \pm 0.1$, $\sigma=0.4$.

\end{itemize}

As expected the spectra of the central components are flatter than
those of the lobes/hotspots. 
The change of the spectral shape
below $\sim$ 1 GHz may be due to either a low-frequency turnover
caused by synchrotron self-absorption, 
or the presence of a spectral break at higher frequencies
caused by the source age. About 70 per cent of the sources have
a peak frequency of the integrated spectrum $\nu_p \geq$100 MHz,
implying a moderate/significant dimming of the flux density at 327 MHz
due to opacity (Table \ref{source}). On the other hand, for 90 per
cent of the objects the spectral
break estimated by \citet{murgia99} is at a frequency below the
highest one used to compute $\alpha_h$,
supporting the steepening produced by radiative losses.\\

\subsection{Notes on individual sources}
\label{section:notes}

Here we provide a brief description of the
sources studied in this paper. 
For sources with a two-sided structure, we provide the flux density
which is computed between the emission from
the whole component (i.e. lobe+hot spot) on each side, unless stated
otherwise. When we have the information on the core position we
computed the arm-length ratio between the lobes.\\

\begin{figure}
\begin{center}
\includegraphics{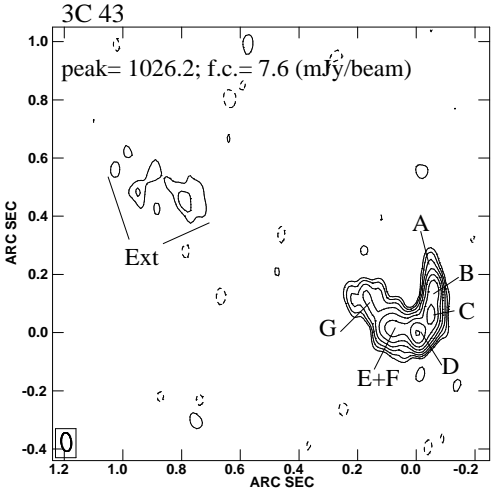}
\vspace{7cm}
\caption{VLBI image at 327 MHz of 3C\,43.}
\label{3c43}
\end{center}
\end{figure}

\begin{figure}
\begin{center}
\includegraphics{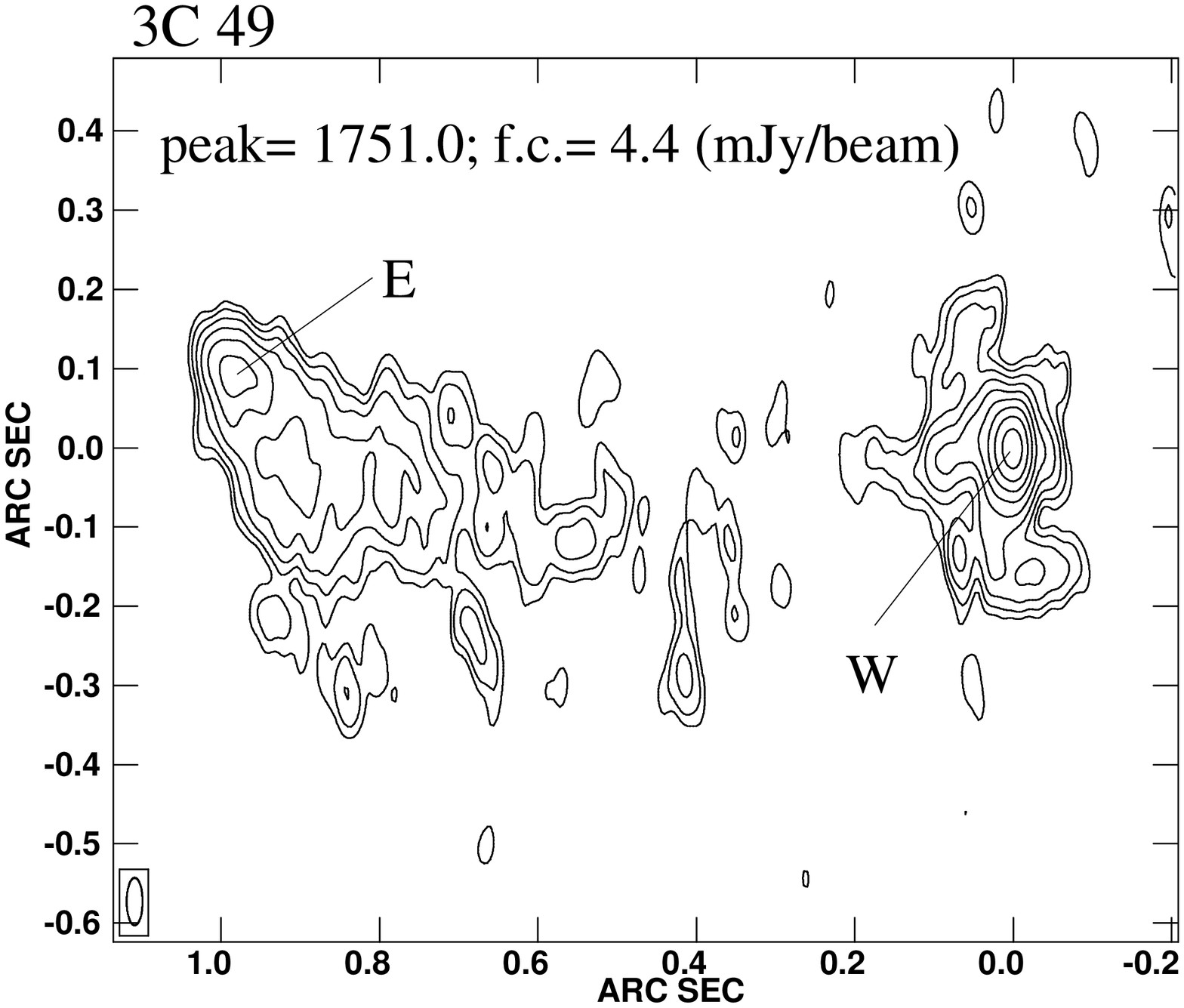}
\vspace{5.5cm}
\caption{VLBI image at 327 MHz of 3C\,49.}
\label{3c49}
\end{center}
\end{figure}

\begin{figure}
\begin{center}
\includegraphics{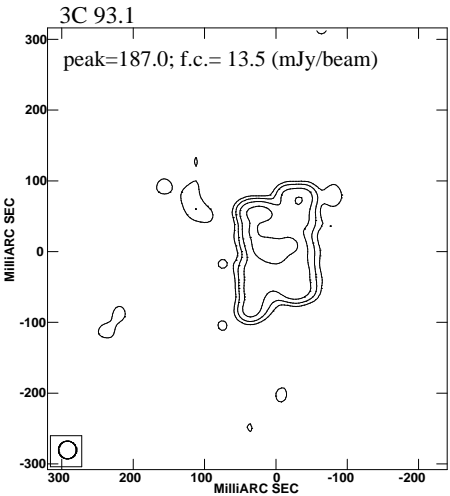}
\vspace{6.8cm}
\caption{VLBI image at 327 MHz of 3C\,93.1.}
\label{3c93}
\end{center}
\end{figure}

\noindent {\bf CSS sources from the 3C sample}\\

\subsubsection{3C\,43~~[~Q, ~z=1.459]}

In our VLBI image at 327 MHz 3C\,43
shows a distorted structure where the radio emission comes mainly from
the western region which is resolved into several sub-components with
a knotty arc-shaped structure
(components from A to G in Fig. \ref{3c43}). 
Component A marks the position of the source
  core that is clearly visible in VLBI images at higher frequencies
  \citep{fanti02}, but it is partially absorbed at 327 MHz.
Only a hint of the
eastern component is detected, while the northern component, imaged at
1.7 GHz with MERLIN observations \citep{fanti02}, is resolved out.\\

\begin{figure}
\begin{center}
\includegraphics{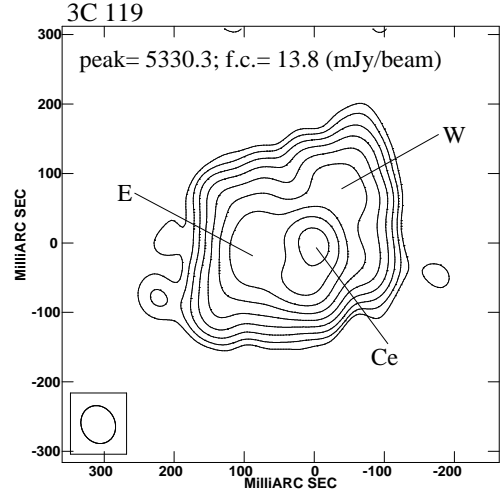}
\vspace{6.5cm}
\caption{VLBI image at 327 MHz of 3C\,119.}
\label{3c119}
\end{center}
\end{figure}

\subsubsection{3C\,49~~[~G, ~z=0.621]}

In our VLBI image at 327 MHz
3C\,49 shows an asymmetric double
structure (Fig. \ref{3c49}) with compact hotspots at each
end.
The spectrum of the eastern lobe is steep, while the spectral 
index of the western component is initially flat
and slowly steepens above 610 MHz (Table \ref{parameter}).
This suggests that the radio emission of this component is dominated
by a compact hotspot, that is likely
self-absorbed
at low frequencies.
The source core, detected at
high frequencies \citep{fanti89,sanghera95,vanbreugel92,ludke98},
is not detected in our observation, and should be considered self-absorbed.
The flux density ratio at 327 MHz is $S_{W}/S_{E}                                
=1.4$, but increases with frequency
owing to
the different spectral indices of the two components.\\
The arm-length ratio is derived from the 5-GHz MERLIN
image \citep{ludke98}, where the source core is clearly
detected, and is $R_{\rm R} = 0.4$, 
where the brighter component is the closer to the
core.\\

\begin{figure}
\begin{center}
\includegraphics{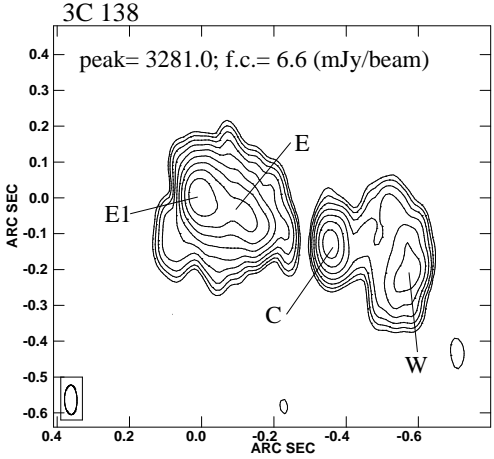}
\vspace{6.0cm}
\caption{VLBI image at 327 MHz of 3C\,138.}
\label{3c138}
\end{center}
\end{figure}

\subsubsection{3C\,93.1~~[~G, ~z=0.243]}

3C\,93.1 has an amorphous radio structure and no compact region is
found in the VLBI image at 327 MHz (Fig. \ref{3c93}).
The spectral index computed between 327 MHz and 1.7 GHz from
\citet{spencer89} is flat with $\alpha=0.0$, but this may be caused by
the severe flux density loss in our VLBI data.\\

\begin{figure}
\begin{center}
\includegraphics{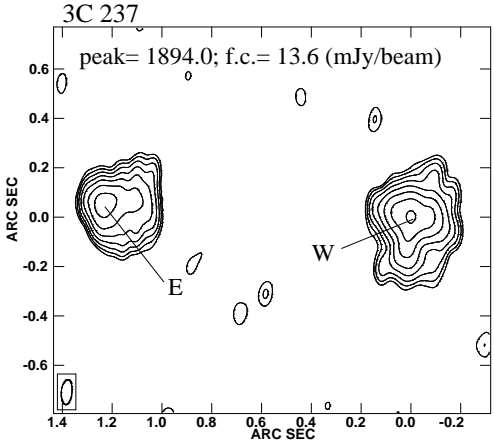}
\vspace{6.0cm}
\caption{VLBI image at 327 MHz of 3C\,237.}
\label{3c237}
\end{center}
\end{figure}

\subsubsection{3C\,119~~[~G, ~z=1.023]}

In our VLBI image at 327 MHz (Fig. \ref{3c119}),
3C\,119 shows a compact component, Ce, slightly elongated 
toward South and surrounded by diffuse emission, in
agreement with data at 610 MHz \citep{nan91}.
The spectral index computed between 327 and 610 MHz is rather flat,
suggesting the presence of unresolved compact components where
self-absorption becomes important, whereas it steepens at higher
frequencies (Table \ref{parameter}).\\

\subsubsection{3C\,138~~[~Q, ~z=0.759]}

The radio source 3C\,138
shows an asymmetric triple structure (Fig. \ref{3c138}). 
The eastern component, E, extending for about 390 mas (2.87 kpc) 
from the centre,
is the brightest one and is interpreted as the approaching
jet \citep{fanti89}. 
Component E1 marks the hotspot. 
The western component, W, is interpreted as the source counter-lobe,
and is located at $\sim$230 mas (1.7 kpc)
from the centre and extends for 440 mas (3.3 kpc) 
in the NS direction.\\
The spectral indices  of the extended components and of the
  hotspot E1, all relatively flat at low frequencies, steepen
above 610 MHz (Table \ref{parameter}). 
The central component, C, has instead an absorbed spectrum, suggesting
that it hosts the source core.
The flux density ratio at 327 MHz
between the eastern and western components is
$S_{\rm E}/S_{\rm W}\sim 8$,
while the arm-length ratio is $R_{\rm R} =1.7$. \\

\begin{figure}
\begin{center}
\includegraphics{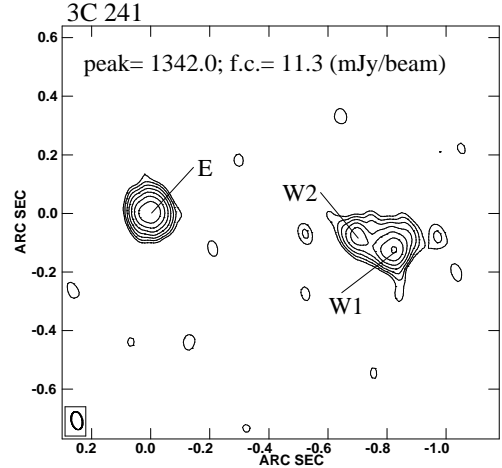}
\vspace{6.5cm}
\caption{VLBI image at 327 MHz of 3C\,241.}
\label{3c241}
\end{center}
\end{figure}

\subsubsection{3C\,237~~[~G, ~z=0.877]}

\noindent The radio source 3C\,237
shows a double morphology in our VLBI image
at 327 MHz (Fig. \ref{3c237}). The eastern component, E, is resolved 
in the EW direction,
while the western component, W, is elongated towards South, in agreement with
what was found in previous works at different frequencies
\citep[e.g.][]{fanti86,nan91,ludke98}. The lobes look edge-brightened,
but no obvious hotspot has been clearly detected at any frequency in the
literature.
The spectra of both lobes E and W have a similar shape, showing a
flattening from high to low frequencies. 
The flux density ratio at 327 MHz
between the western and eastern component is $S_{\rm W}/S_{\rm E}
\sim 1.6$, which remains roughly constant at higher frequencies,
\citep{ludke98},
while the arm-length ratio, measured on the 5-GHz image
where the core is detected, is $R_{\rm R} =1.5$.\\

\begin{figure}
\begin{center}
\includegraphics{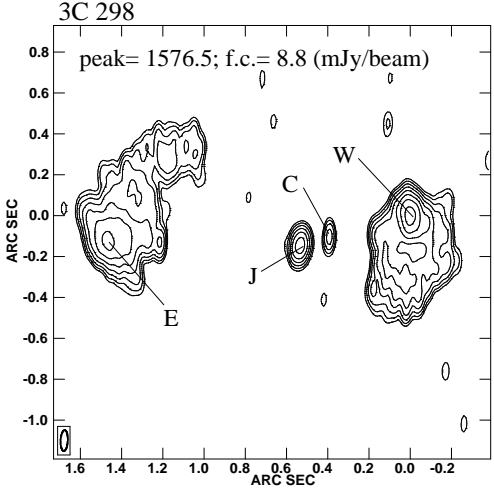}
\vspace{6.5cm}
\caption{VLBI image at 327 MHz of 3C\,298.}
\label{3c298}
\end{center}
\end{figure}

\subsubsection{3C\,241~~[~G, ~z=1.617]}

The radio source 3C\,241 shows a double structure in our VLBI image at 327
MHz (Fig. \ref{3c241}). 
The western component is resolved into two compact regions, W1 and W2,
separated by 120 mas (1.0 kpc)
and aligned with the weak flat-spectrum source core detected by MERLIN
observations at 5 GHz 
\citep{sanghera95} and at 15 GHz \citep{vanbreugel92}.
The spectra of components W1 and W2 are very similar, slightly flat at
low frequencies followed by a steepening with the same slope above 610
MHz. The eastern lobe has a steeper spectrum
at both low and high frequencies (Table \ref{parameter}).
The flux density ratio at 327 MHz
between the eastern (E) and western (W1+W2) component is $S_{\rm E}/S_{\rm W}
\sim$2.3.
Owing to its flatter
spectrum above 610 MHz, 
the western component becomes the brighter at 5 GHz with $S_{\rm W}/S_{\rm E}
\sim 2.2$ \citep{ludke98}. 
Following our assumption that the arm-length ratio is below unity in case of
a brighter-when-closer asymmetry, we have that $R_{\rm R} =1.4$ at 327
MHz, while it switches to 0.7 at 5 GHz (Fig. \ref{asym}).\\

\begin{figure}
\begin{center}
\includegraphics{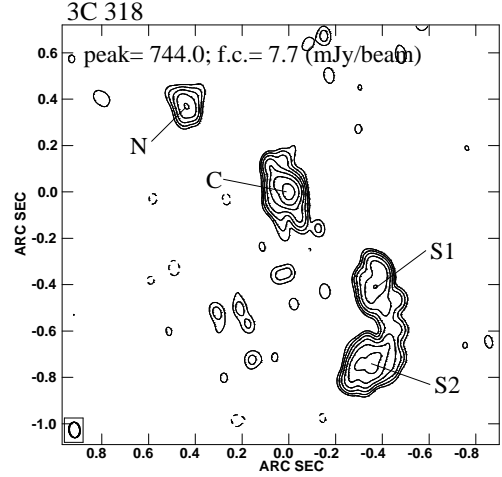}
\vspace{6.5cm}
\caption{VLBI image at 327 MHz of 3C\,318.}
\label{3c318}
\end{center}
\end{figure}

\subsubsection{3C\,298~~[~Q, ~z=1.437]}

In our VLBI image at 327 MHz, 
3C\,298 shows an S-shaped
structure dominated by the emission from the two lobes extending
clockwise with respect to the embedded
compact hotspots (Fig. \ref{3c298}). 
Hotspots are clearly detected
and their distance from the
core is about 930 mas (7.9 kpc) and
560 mas (4.8 kpc) 
for the eastern and western hotspot, respectively. 
A low frequency turnover is visible for the western lobe, likely caused by the
embedded self-absorbed component W. Component E,
instead, has a straight spectrum over the entire frequency range
(Table \ref{parameter}).
Two compact components, J and C, are located
between the lobes. 
They have inverted spectra at frequencies below 610 MHz, and steepen
at higher frequencies (Table \ref{parameter}). 
VLBI images at 1.7 and 5 GHz presented by
\citet{fanti02} show that J is a compact knot, while C is the core.
The flux density ratio between the western and eastern component is
$S_{\rm W} /S_{\rm E} =1.2$ at 327 MHz, and increases
to 1.8 at 5 GHz \citep{fanti02}.
The arm-length ratio is $R_{\rm R} = 0.4$, where the western component is
the closer to the core. \\

\begin{figure}
\begin{center}
\includegraphics{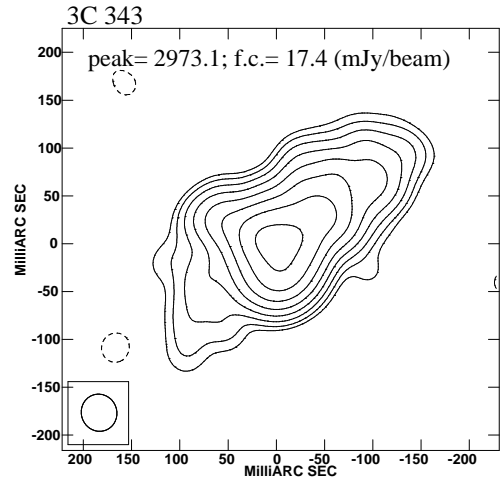}
\vspace{6.5cm}
\caption{VLBI image at 327 MHz of 3C\,343.}
\label{3c343}
\end{center}
\end{figure}

\subsubsection{3C\,318~~[~Q, ~z=1.574]}

In our VLBI image at 327 MHz (Fig. \ref{3c318}),
3C\,318 shows an asymmetric triple structure, where the central
component, labelled C, likely hosts the weak 
source core and is
located at about 0.6 arcsec (5.0 kpc)
and 0.4 arcsec (3.5 kpc) far from the northern and
southern components, respectively.   
A one-sided jet, characterized by a bright knot, emerges from the
north-eastern part of the core region.
The spectral index at low frequencies of N, C, and S1+S2 is
relatively flat (Table \ref{parameter}), but this may be caused by the
significant flux density loss of our VLBI data.
The flux density ratio at 327 MHz
between the southern (S1$+$S2) and northern component is $S_{\rm S}/S_{\rm N}
\sim 9.7$.

\subsubsection{3C\,343~~[~Q, ~z=0.988]}

In our VLBI image at 327 MHz (Fig. \ref{3c343}) 
3C\,343 is elongated in the North-West
direction, in agreement with the structure found by \citet{nan91}
at 610 MHz and by \citet{ludke98} in
5-GHz MERLIN data.
The spectrum of
the whole source increasingly steepens with increasing
frequencies (Table \ref{parameter}).

\begin{figure}
\begin{center}
\includegraphics{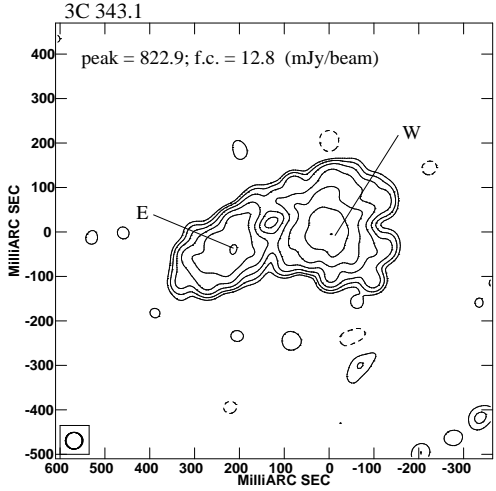}
\vspace{7.0cm}
\caption{VLBI image at 327 MHz of 3C\,343.1.}
\label{3431}
\end{center}
\end{figure}

\subsubsection{3C\,343.1~~[~G, ~z=0.75]}

In our VLBI image at 327 MHz
(Fig. \ref{3431}), 3C\,343.1 is resolved into two
asymmetric lobes separated by 
about 215 mas (1.6 kpc). Component W is slightly elongated in the EW direction,
while component E has a position angle of about 125$^{\circ}$. 
The core has been found neither in our image nor in the literature.
The spectra of the two components are similar: initially quite flat,
they steepen above 610 MHz
(Table \ref{parameter}).
The flux density ratio at 327 and 610 MHz is $S_{\rm W}/S_{\rm E} \sim$2.\\

\begin{figure}
\begin{center}
\includegraphics{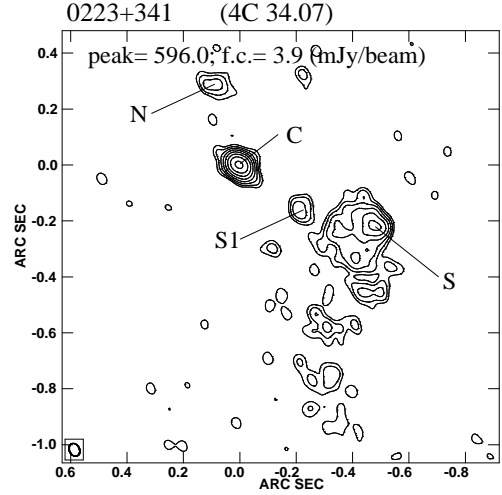}
\vspace{7.0cm}
\caption{VLBI image at 327 MHz of 0223$+$341 (4C\,34.07).}
\label{0223}
\end{center}
\end{figure}

\noindent {\bf CSS and GPS sources from the PW sample}\\

\subsubsection{0223$+$341~~[~Q, ~z=2.91]}

At 327 MHz, 0223$+$341 (alias 4C\,34.07) displays a
complex structure dominated by a bright compact component, labelled C
in Fig. \ref{0223}, which is located between two extended
features, labelled N and S.
Another bright compact structure,
labelled S1 in Fig. \ref{0223} is located at about 0.27 arcsec
(2.1 kpc) from component C in the direction of
the southern lobe. The southern component terminates to the west with
a compact feature, possibly a hotspot. 
The extended emission present
South of component S is clearly visible in VLA images at lower
resolution \citep{lenc08}, but it is almost totally resolved out in our VLBI
image. 
The presence of the northern lobe, detected in deep 327-MHz VLBI data
\citep{lenc08}, 
suggests that component C is likely hosting the source core, as also
proposed by \citet{lenc08}, but also see \citet{mantovani13} for a
different interpretation.
The flux density ratio at 327 MHz between component N and
S is $S_{\rm S}/S_{\rm N} = 17$, while the arm-length ratio is $R_{\rm
  R} = 1.7$ if we assume that component C is the source core.\\

\begin{figure}
\begin{center}
\includegraphics{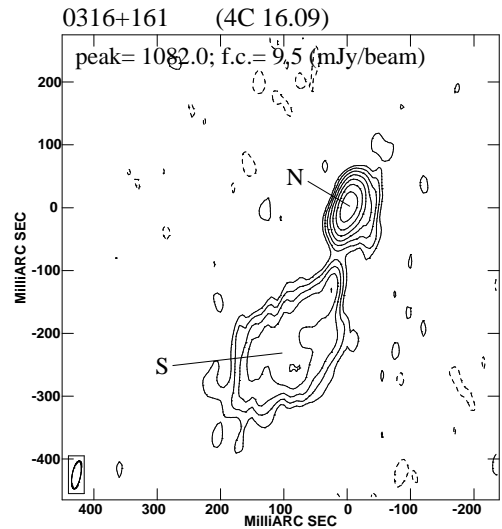}
\vspace{7.0cm}
\caption{VLBI image at 327 MHz of 0316$+$161 (4C\,16.09).}
\label{0316}
\end{center}
\end{figure}

\subsubsection{0316$+$161~~[~G, ~z=0.907]}

At 327 MHz the radio emission of 0316$+$161 (alias 4C\,16.09) is
dominated by a bright elongated component, labelled N in
Fig. \ref{0316}, from which a diffuse low-surface brightness structure
emerges and extends for about
300 mas (2.3 kpc) to the South.
In our image the core would be located roughly at the northern
tip of the extended structure, labelled S in Fig. \ref{0316}. 
The northern component has initially an inverted spectrum
which then switches its trend and steepens, while the souther
component has a steep spectrum.
The flux density ratio at 327 MHz between the southern
and northern component is $S_{\rm S}/S_{\rm N} =1.1$. Owing to its
flat spectrum, the northern component becomes the
brighter at high frequency with
$S_{\rm  N}/S_{\rm S} =14$ at 1.7 GHz \citep{dd13}.
Following our assumption that the arm-length ratio is below unity in case of
a brighter-when-closer asymmetry, we have that $R_{\rm R} =1.7$ at 327
MHz, while it switches to 0.6 at 1.7 GHz (Fig. \ref{asym}).\\

\begin{figure}
\begin{center}
\includegraphics{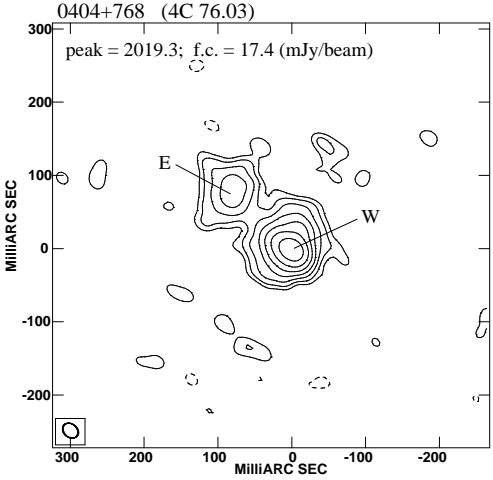}
\vspace{7.0cm}
\caption{VLBI image at 327 MHz of 0404$+$768 (4C\,76.03).}
\label{0404}
\end{center}
\end{figure}

\subsubsection{0404$+$768~~[~G, ~z=0.5985]}

At 327 MHz 0404$+$768 (alias 4C\,76.03) shows an
asymmetric double structure (Fig. \ref{0404}). 
In images at higher resolution \citep{dd95, dd13}
the core is clearly visible and is characterized by an inverted spectrum.
The western lobe has a rather flat, almost straight spectrum from 327
MHz to 8.4 GHz, while
the spectrum of the eastern component is steeper
(Table \ref{parameter}).
The flux density ratio between the western and
eastern component is $S_{\rm
W}/S_{\rm E} =$5.9 at 327 MHz. It becomes 9.5 and 7.0 at 
1.7 and 5 GHz, respectively
\citep{dd13}, owing to the different slope of the spectra.
The arm-length ratio derived from the VLBI image
at 5 GHz is $R_{\rm R} \sim$0.6, with the brighter
component being the closer to the core.\\

\begin{figure}
\begin{center}
\includegraphics{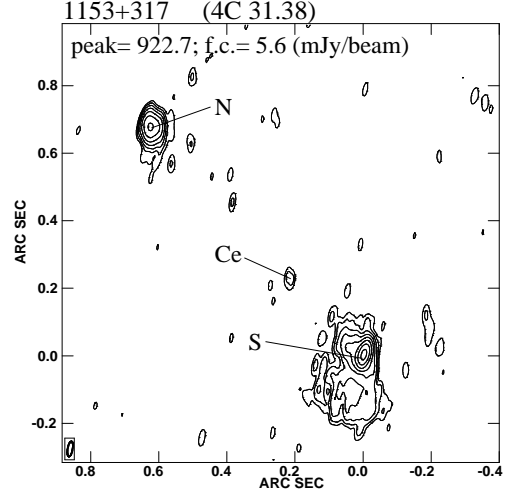}
\vspace{7.0cm}
\caption{VLBI image at 327 MHz of 1153$+$317 (4C\,31.38).}
\label{1153}
\end{center}
\end{figure}

\subsubsection{1153$+$317~~[~Q, ~z=0.417]}

In our VLBI image at
327 MHz, 1153$+$317 (alias 4C\,31.38) shows a very asymmetric double
structure with a weak, possibly compact, component in between
(labelled Ce in Fig. \ref{1153}).
The northern lobe, labelled N, is
slightly resolved, while the southern lobe, labelled S, is characterized by a
bright and compact region, visible only at 327 MHz, that is possibly a
hotspot aligned with components Ce and N. Diffuse emission flows from
the hotspot to the south. 
The spectral information is not adequate to
unambiguously classify Ce component as the source core.
The spectra of the two lobes, quite flat at low frequencies, steepen
above 1.7 GHz (Table \ref{parameter}).
The flux density ratio at 327 MHz is $S_{\rm S}/S_{\rm N}=$2.7, 
becoming smaller ($S_{\rm S}/S_{\rm N}=$2) 
at higher frequencies \citep{spencer89}, as the contribution of the
diffuse emission of the southern component gets less important. \\

\begin{figure}
\begin{center}
\includegraphics{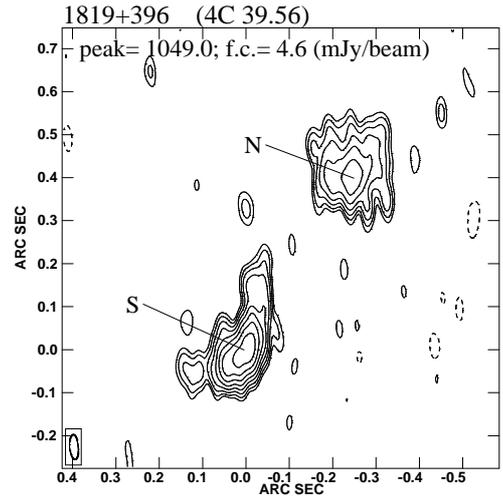}
\vspace{7.0cm}
\caption{VLBI image at 327 MHz of 1819$+$396 (4C\,39.56).}
\label{1819}
\end{center}
\end{figure}

\subsubsection{1819$+$396~~[~G, ~z=0.798]}

In our VLBI image at 327 MHz (Fig. \ref{1819}), 
1819$+$396 (alias 4C\,39.56)
has a double morphology which extends for about
0.7 arcsec (5.3 kpc). The radio emission is dominated by
the southern component, S, whose structure resembles a curved
jet. 
The spectra of both lobes have a break above 1.7 GHz, the southern
component being the flatter at low frequencies (Table \ref{parameter}).
The flux density ratio between component S and N
is $S_{\rm S}/S_{\rm N} = 2.6$. \\

\begin{figure}
\begin{center}
\includegraphics{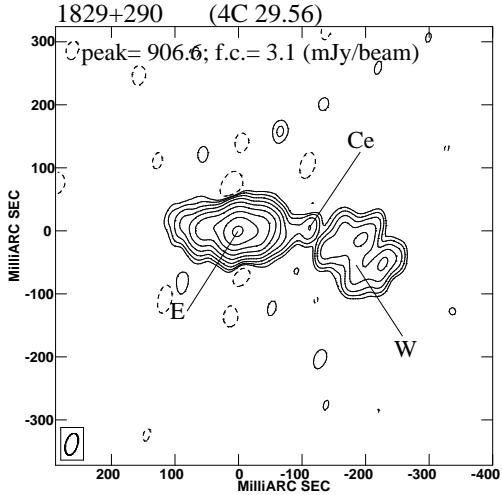}
\vspace{7.0cm}
\caption{VLBI image at 327 MHz of 1829$+$290 (4C\,29.56).}
\label{1829}
\end{center}
\end{figure}

\subsubsection{1829$+$290~~[~G, ~z=0.842]}

In our VLBI image at 327 MHz of 1829$+$290 (alias 4C\,29.56), 
only the central component, labelled E, is detected, while the faint
lobes, visible in MERLIN images \citep{spencer89}, are not detected
(Fig. \ref{1829}).
In addition we observe a central component, Ce, and a weak western
  component, W, undetected in previous VLBI observations at 1.7 GHz
 \citep{dd95}, which is roughly 
oriented in the same direction of the southern lobe. Component Ce does
not seem to have any counterpart at 1.7 GHz. 
The radio spectrum of component E is initially quite flat,
and then curves slowly, becoming straight and steep above 5 GHz.
The flux density ratio between components E and W
is $S_{\rm E}/S_{\rm W} =
7.3$ at 327 MHz. \\

\begin{figure}
\begin{center}
\includegraphics{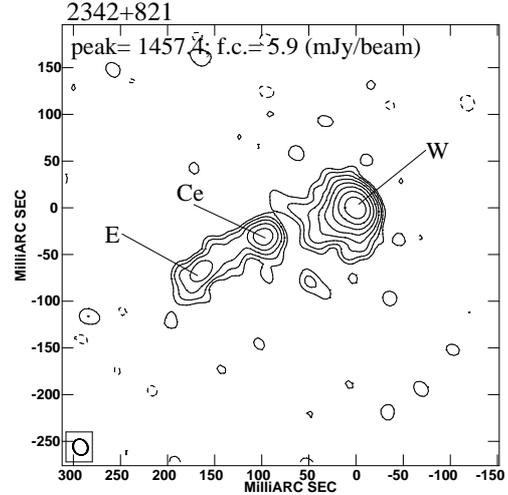}
\vspace{7.0cm}
\caption{VLBI image at 327 MHz of 2342$+$821.}
\label{2342}
\end{center}
\end{figure}

\subsubsection{2342$+$821~~[~Q, ~z=0.735]}

In our VLBI image at 327 MHz, 2342$+$821 displays a triple
well-aligned radio structure 
with a bright component to the west that dominates the radio emission
(Fig. \ref{2342}). A jet-like 
feature is present in the south-east direction.
The initial spectra of components W and Ce are rather flat, and both
steepen above 1.7 GHz. Component E, instead, has an approximately
straight steep spectrum. No evidence of the source core has been found.
The flux density ratio between component W and E
is $S_{\rm W}/S_{\rm E} = 9.5$. \\

\begin{table*}
\caption{Component parameters. Column 1: source
  name; Column 2: source component; Column 3: VLBI flux density at 327
  MHz;
  Columns 4 and 5: major and minor axis; Column 6: position angle of the major axis;
  Column 7:
  spectral index between 327 and 610 MHz, if not stated otherwise; Column
  8: high frequency spectral index. See the notes for the frequency pairs.}
\begin{center}
\begin{tabular}{cccccccc}
\hline
Source&Comp.&$S_{\rm 327}$& $\theta_{\rm maj}$&$\theta_{\rm 
min}$&pa&$\alpha_l$&$\alpha_h$ \\ 
      &     &   Jy     & mas & mas & deg & & \\
(1)&(2)&(3)&(4)&(5)&(6)&(7)&(8) \\ 
\hline
3C\,43    & A+B &   0.72$\pm$0.07 &  47$\pm$1&12$\pm1$& 167$\pm$2 &
1.3 & - \\ 
          & C   &  1.15$\pm$0.11  &  58$\pm$1&20$\pm$1&   166$\pm$1 &
0.3 & - \\ 
          & D   &  1.38$\pm$0.14  &  26$\pm$1&20$\pm$1&  94$\pm$2 &
0.4 &0.6$^{c\,e}$ \\ 
          &E+F  &  1.05$\pm$0.10  & 102$\pm$2 & 28$\pm$1&  90$\pm$1&
1.0 & - \\ 
          & G   &   0.36$\pm$0.04 & 157$\pm$6& 91$\pm$4& 40$\pm$5 &2.7
& - \\ 
          &W$^a$&   0.39$\pm$0.04 & 370&200& 65 & - & - \\
3C\,49    & E   &   0.91$\pm$0.09 &  79$\pm$2 & 52$\pm$2 & 45$\pm$2 &
1.1&1.1$^{c\,g}$ \\ 
          & E$^a$  &  3.07$\pm$0.31 & 800 & 460 & 80 - & 1.2$^b$ &
1.4$^{d\,g}$ \\ 
          & W   &  3.22$\pm$0.03     &  31$\pm$1&28$\pm$1 &13$\pm$1 &
0.0 &1.5$^{d\,e}$ \\ 
          & W$^a$  &  4.31$\pm$0.43     & 420 & 290 & 90 & 0.5$^b$&
0.9$^{d\,g}$ \\ 
3C\,93.1  &Total$^a$&  2.18$\pm$0.22 & 170 & 120 & 10 & - & -\\ 
3C\,119   & E   & 10.12$\pm$1.01  & 130$\pm$1 & 97$\pm$1& 52$\pm$1 &
0.0 & - \\ 
          & Ce  &  5.61$\pm$0.56  &  $<$52 & - & - & 0.0$^b$
&0.8$^{d\,e}$ \\ 
          & W   &  2.23$\pm$0.22 & 137$\pm$1 & 51$\pm$1& 17$\pm$1 & -
& - \\ 
          &Total&  -              &  -         &   -      & -
&0.6$^{k}$&1.9$^{f,h}$ \\ 
3C\,138   & E1  &  6.64$\pm$0.66 &   72$\pm$1 & 51$\pm$1 & 17$\pm$1&
0.0&1.5$^{c\,e}$ \\ 
          & E$^a$  & 16.00$\pm$1.60 & 400 & 340 & 60 &
0.5&0.8$^{d\,h}$ \\ 
          & C   &  1.32$\pm$0.13  &  58$\pm$1 & 42$\pm$1 & 8$\pm$2&
-0.3&0.6$^{c\,e}$ \\ 
          & W$^a$& 1.99$\pm$0.20& 420 & 260 & 175 & 0.2&1.5$^{c\,e}$
\\ 
3C\,237   & E$^a$& 6.55$\pm$0.66 & 400 & 350 & 90 & 0.7$^b$
&1.3$^{d\,g}$ \\ 
          & W$^a$& 10.56$\pm$1.06 & 560 & 420 & - & 0.7$^b$
&1.2$^{d\,g}$ \\ 
3C\,241   & E   &  3.40$\pm$0.34 &  64$\pm$1&45$\pm$1 & 104$\pm$1 &
0.7 &1.8$^{d\,g}$ \\
          & W2  &   0.70$\pm$0.07 &  76$\pm$1& 38$\pm$1 & 66$\pm$2 &
0.5 &1.2$^{c\,g}$ \\ 
          & W1   &  1.80$\pm$0.18 &  63$\pm$1 & 44$\pm$1 & 118$\pm$1&
0.3 &1.2$^{c\,g}$ \\ 
3C\,298   & E   &  5.98$\pm$0.60  & 187$\pm$2&156$\pm$1 & 50$\pm$1 &
1.4$^{b}$ &1.4$^{c\,e}$ \\ 
          & E$^a$&  8.55$\pm$0.85 & 900 & 520 & 120 &
0.7$^b$&1.3$^{d\,h}$ \\ 
          & J   &   0.72$\pm$0.07 &  48$\pm$1& 12$\pm$3 & 107$\pm$1 &
-1.3 &1.0$^{c\,h}$ \\ 
          & C   &   0.11$\pm$0.01 & - & - & - & -0.6$^{b}$& 0.5$^{d\,h}$\\  
          & W   &  4.93$\pm$0.49 & 99$\pm$1 & 71$\pm$1 & 27$\pm$1 &
0.5$^{c}$ & 1.1$^{d\,h}$ \\                                                  
         & W$^{a}$  & 10.73$\pm$1.07 & 680 & 420 & 0 &-  & - \\  
3C\,318   & N & 0.33$\pm$0.03 &  82$\pm$2 &57$\pm$2 & 25$\pm$3 &
0.2$^b$&- \\
          & Ce$^a$& 2.10$\pm$0.21 & 360 & 160 & 30 & 0.0&0.8$^{d\,g}$\\
          & S1$^a$ &  1.19$\pm$0.12 & 330 &200 & 20 & 0.9$^b$& - \\ 
          & S2$^a$ &  1.93$\pm$0.19 & 330 & 220 & 120 & - & - \\ 
          & S1+S2   &   -           & -   &  -  & -
&0.2&1.4$^{d\,h}$ \\ 
3C\,343   &Total$^a$& 12.78$\pm$1.28 & 360 & 180 & 125 & 0.4
&1.0$^{d\,h}$ \\ 
3C\,343.1&E$^a$& 3.73$\pm$0.37&  230 & 180 & 120 & 0.2&1.0$^{c\,h}$\\ 
         &W$^a$& 7.96$\pm$0.80&  300 & 260 & - & 0.4&1.1$^{c\,h}$ \\ 
\hline                               
0223+341  &N    &    0.04$\pm$0.01&  75$\pm$2 & 20$\pm$2 & 80$\pm$1 &-
 & -\\
          &C    &   0.98$\pm$0.10 &  45$\pm$2 & 25$\pm$2 &70$\pm$1 &
-0.6$^b$&0.6$^{d\,h}$ \\ 
          &S1    &  0.08$\pm$0.01   &  70$\pm$2 & 50$\pm$2 &16$\pm$2 &
1.0 &1.0$^{c\,g}$ \\ 
          &S     &  0.44$\pm$0.04  &  70$\pm$2 &60$\pm$1  & 72$\pm$2 &
1.1 &1.1$^{c\,f}$ \\ 
          &S$^a$ &  1.15$\pm$0.11 &   880 &380 & 170 &1.1&
1.1$^{c\,h}$ \\ 
0316+161  &N     &  2.70$\pm$0.27 &  35$\pm$1 &22$\pm$1 & 139$\pm$2 &
-0.5$^b$&1.3$^{d\,g}$ \\
          &S$^a$&  3.19$\pm$0.32 & 320 & 180 & 140 & 1.2$^b$
&2.4$^{d\,e}$ \\ 
0404+768 &E& 1.28$\pm$0.13&   54$\pm$1 & 43$\pm$1 & 7$\pm$2 & 0.7$^b$
&0.9$^{d\,e}$ \\ 
         &W& 7.54$\pm$0.75& 36$\pm$1 & 31$\pm$1 & 41$\pm$1 & 0.3$^b$
&0.5$^{d\,e}$ \\ 
1153+317  &N     &  1.47$\pm$0.15 &  43$\pm$1 &33$\pm$1 & 44$\pm$2 &
0.3$^b$ &1.1$^{d\,g}$ \\ 
          &Ce    &    0.03$\pm$0.01    &  - & - & - &0.5$^b$
&0.5$^{c\,g}$ \\ 
          &S     &  2.16$\pm$0.22    &  34$\pm$2& 24$\pm$2 &
145$\pm$1& - & -\\
          &S$^a$&  3.95$\pm$0.40 & 410 & 140 & 150 & 0.5$^b$
&1.1$^{d\,g}$ \\ 
\hline
\end{tabular}
\end{center}
$^a${The flux density was derived by \texttt{TVSTAT} in
  \texttt{AIPS}. The component angular size was measured on the
  contour image and is roughly twice the FWHM of a conventional
  Gaussian covering a similar area.}\\
$^b${Spectral index computed between 327 MHz and 1.7 GHz.
Two letters indicate other frequency pairs:$^c$=0.610, $^d$=1.7 GHz,
$^e$=5 GHz, $^f=$8.4 GHz, $^g$=15 GHz, $^h$=22~GHz, $^k~\alpha_{0.327}^{8.4}$.}
\label{parameter}
\end{table*}

\addtocounter{table}{-1}
\begin{table*}
\caption{Continued.}
\begin{center}
\begin{tabular}{ccrccccc}
\hline
Source&Comp.&$S_{\rm 327}$& $\theta_{\rm maj}$&$\theta_{\rm
  min}$&pa&$\alpha_l$&$\alpha_h$ \\
      &     &   Jy     & mas & mas & deg & & \\ 
(1)&(2)&(3)&(4)&(5)&(6)&(7)&(8) \\ 
\hline
1819+396  &N$^a$&  1.55$\pm$0.16 & 240 & 190 &100 & 0.8$^b$&
1.4$^{d\,g}$ \\ 
          &S$^a$&  4.24$\pm$0.42 & 370 & 220 & 130 & 0.4$^b$
&1.1$^{d\,g}$ \\ 
1829+290  &W$^a$&   0.45$\pm$0.04  & 180 & 150 & 40 & - & - \\ 
          &Ce    &    0.03$\pm$0.01    & -   &-  &- & -  & - \\ 
          &E     &  2.93$\pm$0.29    &  54$\pm$1 & 24$\pm$1 &
90$\pm$2& - & -\\
          &E$^a$&  3.27$\pm$0.32 & 260 & 170 & 180 &0.2$^b$  &
1.2$^{e\,h}$ \\
2342+821  &W     &  3.32$\pm$0.33  &  21$\pm$2 &18$\pm$2 &42$\pm$1 & -
&- \\ 
          &W$^a$&  3.71$\pm$0.37 & 100 &90 &150  &0.2$^b$& 0.8$^{d\,e}$
\\ 
          &Ce    &   0.31$\pm$0.03 &  20$\pm$2 & 12$\pm$1 & 103$\pm$1
& 0.4$^b$ &1.0$^{d\,e}$ \\ 
          &E     &   0.35$\pm$0.04 &  60$\pm$2 & 15$\pm$1 & 129$\pm$1
& 0.9$^b$&1.0$^{d\,e}$ \\ 
\hline
\end{tabular}
\end{center}
\end{table*}

\begin{figure}
\begin{center}
\includegraphics{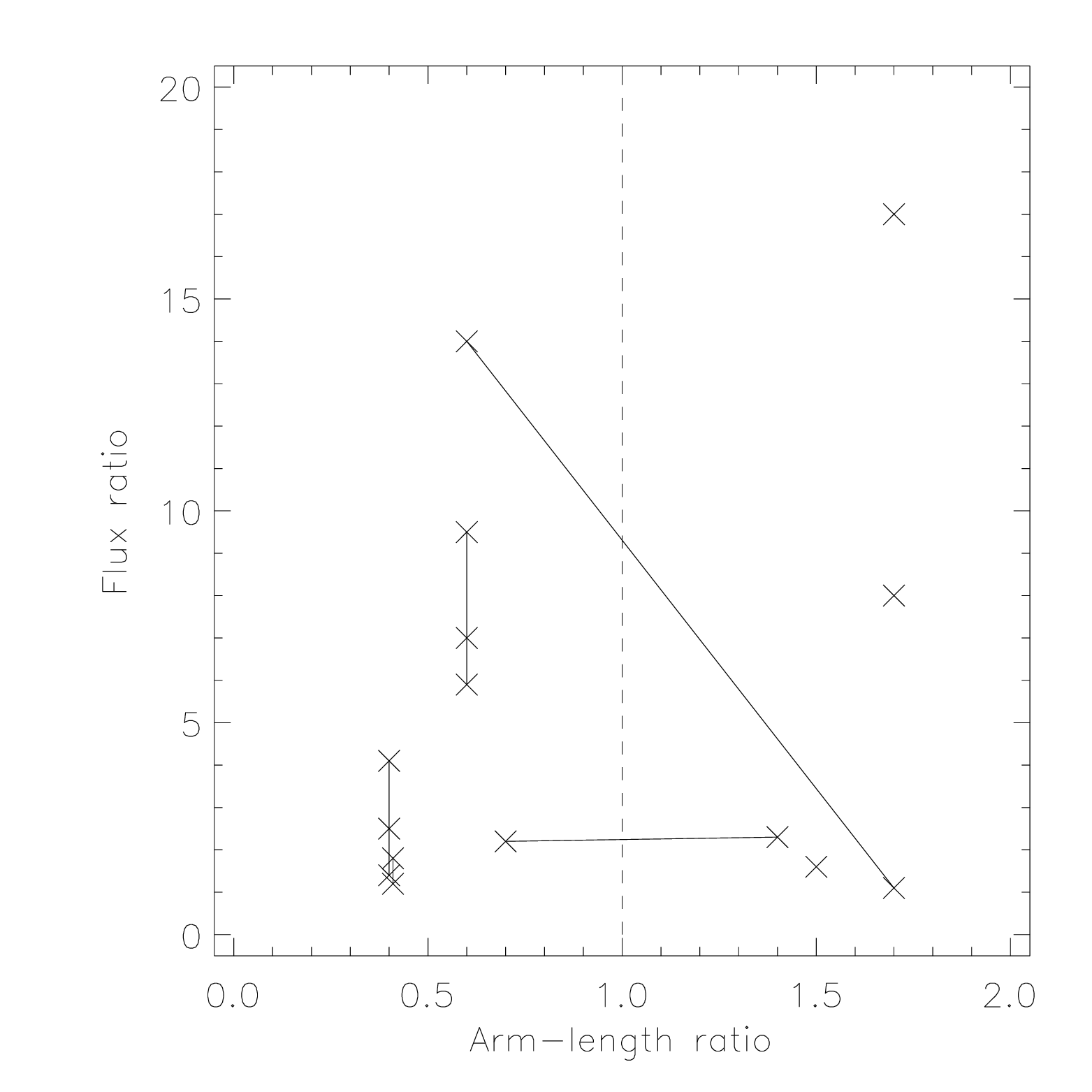}
\vspace{8.5cm}
\caption{Flux density ratio vs arm-length ratio for the 8 CSS/GPS
  sources with a core detection. The vertical dashed line corresponds
  to an arm-length ratio of unity. Values below the unity indicates
  that the brighter component is the closer to the core. When the flux
  density could be measured at various frequencies, points are
  connected with a line. A change in the arm-length ratio for the same
source indicates that the brighter component at low frequency becomes
the fainter at high frequency.}
\label{asym}
\end{center}
\end{figure}

\section{Discussion}
\label{discussion}

Diffuse low-surface brightness regions are difficult to observe in
VLBI observations at centimeter wavelengths due to their steep spectra and
large angular scale. These features mark the place where reside the
oldest electron populations that radiate at low frequencies as a
consequence of the severe energy losses they experienced during their
lifetime. This is particularly true for compact lobes, like
those of CSS/GPS sources, where the magnetic field is substantially
stronger than that in the lobes of classical Fanaroff-Riley radio
galaxies \citep[e.g.][]{fanti95,croston05}. \\
Our low-frequency observations are usually dominated by low-surface
brightness diffuse emission, whereas compact components, like hotspots
and cores becomes predominant at higher frequencies. This is reflected
in the analysis of the flux density and arm-length asymmetry. For the
8 sources with an unambiguous core detection either in our images or in
the literature, we find that among the four with $R>2$ (i.e. a
significant asymmetry in flux density) in only one of them the brighter
component is the closer to the core. This low fraction is in contrast
with the much higher percentage ($\sim$50 per cent) that is found at
higher frequencies \citep{dd13}. However, if we consider the flux
density ratio at higher frequencies also for our target sources, the
situation changes and the percentage of sources with a brighter-when-closer
behaviour increases to
50 per cent (Fig. \ref{asym}). This is caused by the different
spectral index of the source components that makes lobe dominate at
low frequencies, while they are overwhelmed by the emission from
hotspots at high frequencies.

\indent Although the surface brightness of lobes is generally lower than that in
cores, jets and hotspots, these extended
structures store a large fraction of the source energy
budget. \\
Given their large volumes, the total energy derived for lobes
(under minimum energy condition \citep{pacho70}; see Appendix \ref{appendix})
is at least one order of magnitude
larger than that derived for the hotposts (Table \ref{phys_tab}).\\
\indent For the sources where it is possible to disentangle the hotspot
contribution from the lobe emission we estimate the source age following
energy budget arguments. The dynamical source age is estimated by:

\begin{equation}
  \tau_{\rm dyn} \sim 2\times E_{\rm min}/F_{\rm e},
  \label{dynamic-age}
  \end{equation}
  
 \noindent where $E_{\rm min}$
is the total energy stored in the lobe, $F_{\rm e}$ is the jet energy flux,
and the factor 2 accounts for the work spent in inflating the
lobe \citep{fanti02}. The jet energy flux is $F_{\rm e} = c \Pi$,
where $\Pi$, the jet thrust, corresponds to
$\Pi = p_{\rm eq}\times A$, with $p_{\rm eq}$ is the pressure in the
hotspot and $A$ is the impact area. We obtain dynamical ages between 
2$\times$10$^{3}$ and 5$\times$10$^{4}$ yr (Table \ref{ages}).
The values derived
in this way should be representative of the order of 
magnitude of the dynamical ages, owing to the strong assumptions on the
minimum energy condition, a constant jet thrust during the source
lifetime, and the large uncertainty on the
volume. On the other hand, the missing flux density should not
influence the estimate more than a factor of 1.3.   
These values are
in agreement with the radiative ages derived by \citet{murgia99}
by fitting the integrated radio spectrum, with some
exceptions. However, if we compare the equipartition magnetic field
that we estimate in the lobes and the one assumed by \citet{murgia99}
we see that the latter is much higher in a few sources, and is more similar to the one
we estimate in the hotspots. A remarkable example is 3C\,49 where
$\tau_{dyn}$ and $\tau_{rad}$ differ more than an order of
magnitude. Assuming their $\tau_{\rm rad}$, \citet{murgia99} estimated
that 3C\,49 is expanding at a superluminal velocity ($v>4c$), which is
unlikely for a radio source, hosted by a galaxy, whose radio emission should not
be affected by beaming effects. If we compute $\tau_{rad}$ assuming the magnetic field in
the lobes we obtain $\tau_{rad} \sim$4$\times$10$^{4}$ yr, which is
similar to $\tau_{\rm dyn}$, while the velocity expansion
becomes about 0.3$c$, consistent with what is found by kinematic
studies of CSO sources \citep[e.g.][]{giroletti09}.\\

\begin{table}
\caption{Physical parameters for the hotspots and lobes.}
\begin{center}
\begin{tabular}{ccc}
\hline
   & Hotspots & Lobes \\
\hline
$L$ (erg s$^{-1}$) & 5$\times$10$^{43}$ - 3$\times$10$^{45}$& 10$^{44}$ - 6$\times$10$^{45}$\\
$E_{\rm min}$ (erg) & 10$^{56}$ - 10$^{58}$& 10$^{57}$ - 6$\times$10$^{59}$\\
$u_{\rm min}$ (erg cm$^{-3}$) & 4$\times$10$^{-8}$ - 10$^{-6}$& 10$^{-9}$ - 3$\times$10$^{-7}$\\
$p_{\rm eq}$ (dyne cm$^{-2}$) & 3$\times$10$^{-8}$ - 6$\times$10$^{-7}$& 2$\times$10$^{-10}$ - 2$\times$10$^{-7}$\\
$H_{\rm eq}$ (mG) & 0.3 - 1.5 & 0.1 - 0.8\\
\hline
\end{tabular}
\end{center}
\label{phys_tab}
\end{table}

\begin{table}
  \caption{Source ages and magnetic fields. Column 1: source name;
    column 2: source component; column 3: dynamical age; column 4:
    equipartition magnetic field of the lobe; columns 5 and 6:
    radiative ages and equipartition magnetic field from
    \citet{murgia99}.}
  \begin{center}
    \begin{tabular}{cccccc}
\hline
      Source & comp. & $\tau_{\rm dyn}$&$H_{\rm eq, lobe}$&$\tau_{\rm
        rad}$&$H_{\rm eq}$\\
      & &10$^4$ yr & mG & 10$^4$ yr & mG\\
\hline
      3C\,49 & W & 2.4 & 0.2 & 0.1 & 7.0\\
             & E & 3.6 & 0.2 & 0.1 & 7.0\\
      3C\,138& E & 3.0 & 0.4 & 1.7 & 1.0\\
      3C\,298& W & 4.8 & 0.4 & $>$5& 1.6\\
             & E & 3.3 & 0.2 & $>$5& 1.6\\
      1153$+$317&S& 0.9 & 0.4 & 0.5& 1.7\\
      2342$+$821&W& 0.2 & 0.5 & $\leq$0.13& 4.5\\
      \hline
    \end{tabular}
  \end{center}
  \label{ages}
  \end{table}

\section{Summary}

We presented VLBI images at 327 MHz of 18 CSS/GPS radio sources from
the 3C and Peacock \& Wall catalogues. The conclusions that we can
draw from this investigation are as follows:

\begin{itemize}

\item All the sources are clearly detected and imaged.
  About 80 per cent of the sources have
    a ``Double/Triple'' morphology,
    and the source core is detected in only 16 per cent of them. This
    is likely the consequence of self-absorption that is severe at the
    observing frequency.

  \item For the 8 sources with information on the core position, either from
    these data or from literature, we studied the asymmetries in flux
    density and arm-length. Half of the sources have a roughly
    symmetry in flux density between the
    two sides of the radio emission. Only in one out of the 4
    asymmetric sources the
    brighter component is the closer to the core. The situation changes if we
consider flux density ratio at higher frequencies, 
where the emission is dominated by compact hotspot components with a
flatter spectrum than the lobes.

\item The dynamical ages derived on energy budget arguments are 
between 2$\times$10$^{3}$ and 5$\times$10$^{4}$ yr, in rough agreement
  with the radiative ages estimated by the analysis of the integrated
  radio spectrum. The larger discrepancy between radiative and
  dynamical ages are found in sources where the integrated spectrum
  and the magnetic field are dominated by the hotspot components. In
  this case the radiative age
  likely represents the time spent by the particles in these regions
  rather than the genuine age of the source.

  \end{itemize}

A comprehensive study of the physics of extragalactic radio sources
requires information covering a large part of the radio spectrum: high
frequencies for picking up the region of fresh particle injection and
acceleration, and low frequencies to infer the energetics of the
radio source. The combination of future facilities, like
next-generation very large array, and LOFAR with the addition of the
international baselines, will be pivotal for population studies of
CSS/GPS sources.\\

\section*{Acknowledgments}

We thank the anonymous referee for reading the manuscript carefully
and making valuable suggestions.
We thank the European VLBI Network and the United States VLBI Network
for carrying out the observations. The WSRT is operated by the
Netherlands Foundation for Research in Astronomy with the financial
support of the Netherlands Organization for Scientific Research (NWO). 
The National Radio Astronomy Observatory is operated by Associated
Universities, Inc., under contract with the National Science
Foundation. This research has made use of the NASA/IPAC Extragalactic
Database (NED), which is operated by the Jet Propulsion Laboratory,
California Institute of Technology, under contract with the National
Aeronautics and Space Administration. DD still acknowledges the Commission
of the European Union for the award of a Fellowship, although it
happened a long ago. The authors thanks A. Medici and Z. Cai for
help in the data reduction.
The WENSS project was a collaboration between the Netherlands 
Foundation for Research in Astronomy and the Leiden Observatory. The
WENSS team consisted of Ger de Bruyn, Yuan Tang, Roeland Rengelink, 
George Miley, Huub R\"ottgering, Malcolm Bremer, Martin Bremer, Wim
Brouw, Ernst Raimond and David Fullagar.

\section*{Data Availability}
Final images are available upon reasonable request.

\appendix
\section{Phisycal parameters}
\label{appendix}

We computed the minimum total energy, $E_{\rm min}$,
the minimum energy density, $u_{\rm min}$, the equipartition pressure
$p_{\rm eq}$,
and the equipartition
magnetic field,$H_{\rm eq}$, using:

\begin{equation}
E_{\rm min} = c_{13} L^{4/7} V^{3/7}
\label{energy}
\end{equation}

\begin{equation}
u_{\rm min} = c_{13} \left( \frac{L}{V} \right)^{4/7}
\label{density}
\end{equation}

\begin{equation}
p_{\rm eq} = \frac{13}{21} u_{\rm min}
\label{pressure}
\end{equation}

\begin{equation}
H_{\rm eq} = \left( \frac{c_{12} L}{V} \right)^{\frac{2}{7}}
\label{magnetic}
\end{equation}

\noindent where $L$ is the radio luminosity, $V$ is the volume
homogeneously filled with relativistic plasma in which electrons and
positrons have identical energy, and
          $c_{12}$ and $c_{13}$ are constants tabulated in
\citet{pacho70} and depend on the spectral index and
          the upper and lower cutoff frequencies. We assumed an
          average
          spectral index $\alpha                                                  
          =0.7$, a lower cutoff frequency, $\nu_{1}$ = 10 MHz, and an
          upper cutoff frequency, $\nu_{2}$ = 100 GHz. The radio
          luminosity $L$ was calculated by:

\begin{equation}
L = \frac{4 \pi D_{\rm L}^2}{(1+z)^{1 - \alpha}}
\int_{\nu_{1}}^{\nu_{2}} S(\nu) d \nu
\label{luminosity}
\end{equation}

\noindent where $D_{\rm L}$ is the luminosity distance and $z$ is the
redshift.
We approximated the volume of the source components
to a prolate ellipsoid:

\begin{equation}
V = \frac{\pi}{6} \left( \frac{D_{\rm L}}{(1+z)^2} \right)^3
\theta_{\rm maj} \theta_{\rm min}^2
\label{volume}
\end{equation}

\noindent where $\theta_{\rm maj}$ and $\theta_{\rm min}$ are the
major
and minor angular size, respectively. \\


\begin{thebibliography}{}

\bibitem[Alexander(2000)]{alexander00}
 Alexander A. 2000, MNRAS, 319, 8

\bibitem[Altschuler et al.(1995)]{altschuler95}
Altschuler, D.R., Gurvits, L.I., Alef, W., Dennison, B., Graham, D.,
  Trotter, A.S., Carson, J.E. 1995, A\&AS, 114, 197
 
\bibitem[An \& Baan(2012b)]{an12}
An, T., Baan, W. A. 2012b, ApJ, 760, 77

\bibitem[An et al.(2012a)]{an12a}
An, T., et al. 2012a, ApJS, 198, 5

\bibitem[Chuprikov et al.(1999)]{chuprikov99}
Chuprikov, A.A., et al. 1999, NewAR, 43, 747
  
\bibitem[Croston et al.(2005)]{croston05}
  Croston J. H., Hardcastle M. J., Harris D. E., Belsole E., Birkinshaw M.,
Worrall D. M., 2005, ApJ, 626, 733

\bibitem[Dallacasa et al.(1995)]{dd95}
Dallacasa, D., Fanti, C., Fanti, R., Schilizzi, R.T., Spencer,
R.E. 1995, A\&A, 295, 27

\bibitem[Dallacasa et al.(2013)]{dd13}
Dallacasa, D., Orienti, M., Fanti, C., Fanti, R., Stanghellini,
C. 2013, MNRAS, 433, 147

\bibitem[Douglas et al.(1996)]{douglas96}
Douglas, J.M., Bash, F.N., Bozyan, F.A., Torrence, G.W., Wolfe,
C. 1996, AJ, 111, 1945

\bibitem[Fanti et al.(1986)]{fanti86}
Fanti, C., Fanti, R., Schilizzi, R.T., Spencer, R.E., van Breugel,
W.J.M. 1986, A\&A, 170, 10

\bibitem[Fanti et al.(1989)]{fanti89} 
Fanti, C., et al. 1989, A\&A, 217, 44

\bibitem[Fanti et al.(1990)]{fanti90}
Fanti, R., Fanti, C., Schilizzi, R.T., Spencer, R.E., Nan R., Parma,
P., van Breugel, W.J.M., Venturi, T. 1990, A\&A, 231, 333 
 
\bibitem[Fanti et al.(1995)]{fanti95}
Fanti, C., Fanti, R., Dallacasa, D., Schilizzi, R.T., Spencer, R.E.,
Stanghellini, C. 1995, A\&A, 302, 317 

\bibitem[Fanti et al.(2000)]{fanti00}
Fanti, C., Pozzi, F., Fanti, R., et al. 2000, A\&A, 358, 499

\bibitem[Fanti et al.(2002)]{fanti02}
Fanti, C., Fanti, R., Dallacasa, D., McDonald, A., Schilizzi, R.T.,
Spencer, R.E. 2002, A\&A, 396, 801

\bibitem[Giroletti \& Polatidis(2009)]{giroletti09}
Giroletti, M., Polatidis, A. 2009, AN, 330, 193

\bibitem[Gugliucci et al.(2005)]{gugliucci05}
Gugliucci, N.E., Taylor, G.B., Peck, A.B., Giroletti, M. 2005, ApJ, 622, 136

\bibitem[Jeyakumar et al.(2000)]{jeyakumar00}
Jeyakumar, S., Saikia, D.J., Pramesh, R.A., Balasubramanian, V. 2000,
A\&A, 362, 27

\bibitem[K\"uhr et al. (1981)]{kuhr81}
K\"uhr, H, Witzel, A., Pauliny-Toth, I.I.K., Nauber, U. 1981, A\&AS,
45, 367 

\bibitem[Labiano et al.(2006)]{labiano06}
Labiano, A., Vermeulen, R.C., Barthel, P.D., O'Dea, C.P.,
Gallimore, J.F., Baum, S., de Vries, W. 2006. A\&A, 447, 481

\bibitem[Laing et al. (1983)]{laing83}
Laing, R.A., Riley, J.M., Longair, M.S. 1983, MNRAS, 204, 151

\bibitem[Lenc et al. (2008)]{lenc08}
Lenc, E., Garrett, M.A., Wucknitz, O., Anderson, J.M., Tingay,
S.J. 2008, ApJ, 673, 78

\bibitem[L\"udke et al. (1998)]{ludke98}
L\"udke, E., Garrington, S.T., Spencer, R.E., Akujor, C.E., Muxlow,
T.W.B., Sanghera, H.S., Fanti, C. 1998, MNRAS, 299, 467 

\bibitem[Mantovani et al.(2013)]{mantovani13}
Mantovani, F., Rossetti, A., Junor, W., Saikia, D.J., Salter,
C.J. 2013, A\&A, 555,4 

\bibitem[Morganti et al.(2013)]{morganti13}
Morganti, R., Fogasy, J., Paragi, Z., Oosterloo, T.,
Orienti, M. 2013, Science, 341, 1082

\bibitem[Murgia et al.(1999)]{murgia99}
Murgia, M., Fanti, C., Fanti, R., Gregorini, L., Klein, U., Mack,
K.-H., Vigotti, M. 1999, A\&A, 345, 769 

\bibitem[Murgia (2003)]{murgia03}
Murgia, M. 2003, PASA, 20, 19

\bibitem[Murphy et al. (2010)]{murphy10}
Murphy, T., et al. 2010, MNRAS, 402, 2403

\bibitem[Nan et al. (1991a)]{nan91}
Nan R., Schilizzi, R.T., Fanti, C., Fanti, R. 1991, A\&A, 252, 527

\bibitem[O'Dea(1998)]{odea98}
O'Dea, C.P. 1998, PASP, 110, 493

\bibitem[O'Dea \& Saikia(2020)]{odea20}
O'Dea, C.P., Saikia, D.J. 2020, arXiv:2009.02750

\bibitem[Orienti(2016)]{mo16}
Orienti, M. 2016, AN, 337, 9 

\bibitem[Pacholczyk(1970)]{pacho70}
Pacholczyk, A. G. 1970, Radio Astrophysics (San Francisco: Freeman \& Co.)

\bibitem[Peacock \& Wall (1981)]{pw81}
Peacock, J.A., Wall, J.V. 1981, MNRAS, 194, 331

\bibitem[Polatidis \& Conway (2003)]{polatidis03}
Polatidis, A.G., Conway, J.E. 2003, PASA, 20, 69 

\bibitem[Rampadarath et al.(2009)]{rampadarath09}
Rampadarath, H., Garrett, M.A., Polatidis, A. 2009, A\&A, 500, 1327
  
\bibitem[Readhead (1994)]{readhead94}
Readhead, A.C.S. 1994, ApJ, 426, 51

\bibitem[Readhead et al.(1996)]{readhead96}
Readhead, A.C.S., Taylor, G.B., Pearson, T.J., Wilkinson, P.N. 1996,
ApJ, 460, 634
  
\bibitem[Rengelink et al.(1997)]{rengelink97}
Rengelink, R.B., Tang, Y., de Bruyn, A.G., Miley, G.K., Bremer, M.N.,
R\"ottgering, H.J.A., Bremer, M.A.R. 1997, A\&AS, 124, 259

\bibitem[Sanghera et al. (1995)]{sanghera95}
Sanghera, H.S., Saikia, D.,J., L\"udke, E., Spencer, R.E., Foulsham,
P.A., Akujor, C.E., Tzioumis, A.K. 1995, A\&A, 295, 629 

\bibitem[Siemiginowska et al. (2005)]{siemiginowska05}
Siemiginowska, A., Cheung, C.C., LaMassa, S., et al. 2005, ApJ, 632, 110

\bibitem[Snellen et al. (2000)]{snellen00}
Snellen I.A.G., Schilizzi R.T., Miley G.K., de Bruyn, A.G., Bremer,
M.N. R\"ottgering, H.J.A. 2000, MNRAS, 319, 445 

\bibitem[Sobolewska et al.(2019)]{sobolewska19}
Sobolewska, M., Siemiginowska, A., Guainazzi, M.,
Hardcastle, M., Migliori, G., Ostorero, L., Stawarz, \L. 2019, ApJ,
871, 71

\bibitem[Spencer et al. (1989)]{spencer89}
Spencer, R.E., McDowell, J.C., Charlesworth, M., Fanti, C., Parma, P.,
Peacock, J.A. 1989, MNRAS, 240, 657

\bibitem[Torniainen et al.(2007)]{torniainen07}
Torniainen, I., Tornikoski, M., L\"ahteenm\"aki, A., Aller, M.F., Aller,
H.D., Mingaliev, M.G. 2007, A\&A, 469, 451

\bibitem[van Breugel et al. (1984)]{vanbreugel84}
van Breugel, W., Miley,G., Heckman, T. 1984, AJ, 89, 5

\bibitem[van Breugel et al. (1992)]{vanbreugel92}
van Breugel, W.J.M., Fanti, C., Fanti, R., Stanghellini, C.,
Schilizzi, R.T., Spencer, R.E. 1992, A\&A, 256, 56 

\bibitem[Zovaro et al.(2019)]{zovaro19}
Zovaro, H.R.M., Sharp, R., Nesvadba, N.P.H., Bicknell,
G.V., Mukherjee, D., Wagner, A.Y., Groves, B.,
Krishna, S. 2019, MNRAS, 484, 3393

\end{thebibliography}
\end{document}